\documentclass[aps,amssymb,amsmath,prl,reprint,noshowpacs]{revtex4-1}
\usepackage{times}
\usepackage{amssymb,amsmath,graphicx}
\usepackage[usenames]{color}
\usepackage{bbold}
\usepackage{siunitx}
\usepackage{braket}

\usepackage[T1]{fontenc}
\usepackage[utf8]{inputenc}
\usepackage[english]{babel}

\usepackage{hyperref}
\hypersetup{colorlinks=true, citecolor=blue, urlcolor=blue, linkcolor=blue}

\usepackage[normalem]{ulem}
\usepackage{lineno}
\newcommand{\NA}{\mathrm{N \mskip-0.5\thinmuskip A}}

\newcommand{\diff}{\text{d}}
\usepackage{textcomp}
\newcommand{\figwidth}{0.95\columnwidth} %% specifies size for pictures
\newcommand{\commentOut}[1]{}
\usepackage{scalefnt}   % to scale font size

\newcommand{\affil}{Photonics Laboratory, ETH Zürich, CH-8093 Zürich, Switzerland}

\begin{document}
\scalefont{1.05}
\title{Motional Sideband Asymmetry of a Nanoparticle Optically Levitated in Free Space}
\author{Felix Tebbenjohanns}
\affiliation{\affil}
\author{Martin Frimmer}
\affiliation{\affil}
%\email{frimmerm@ethz.ch}
\homepage{http://www.photonics.ethz.ch}
\author{Vijay Jain}
\affiliation{\affil}
\author{Dominik Windey}
\affiliation{\affil}
%\altaffiliation[Present address: ]{IBM Research-Zurich, Säumerstrasse 4, CH-8803 Rüschlikon, Switzerland}
\author{Lukas Novotny}
\affiliation{\affil}

% Reviewer suggestions: 

\begin{abstract}
The hallmark of quantum physics is Planck's constant $h$, whose finite value entails the quantization that gave the theory its name. 
The finite value of $h$ gives rise to inevitable zero-point fluctuations even at vanishing temperature. 
The zero-point fluctuation of mechanical motion becomes smaller with growing mass of an object, making it challenging to observe at macroscopic scales. Here, we transition a dielectric particle with a diameter of 136~nm from the classical realm to the regime where its zero-point motion emerges as a sizeable contribution to its energy. To this end, we optically trap the particle at ambient temperature in ultrahigh vacuum and apply active feedback cooling to its center-of-mass motion.
We measure an asymmetry between the Stokes and anti-Stokes sidebands of photons scattered by the levitated particle, which is a signature of the particle's quantum ground state of motion.
\end{abstract}

\date\today

\maketitle

\paragraph{Introduction.}
A paradigm of quantum mechanics is a mass bound in a harmonic potential with angular oscillation frequency $\Omega$. According to quantum theory, the state of the mass can be described as a superposition of energy eigenstates. These states are enumerated by the quantum (or occupation) number $n$ with respective energies $E_n=\hbar\Omega(n+1/2)$, where $\hbar=h/(2\pi)$ is the reduced Planck constant~\cite{Planck1910,Heisenberg1925,Cohen-Tannoudji1991}. For a harmonic oscillator coupled to a thermal bath at temperature $T$, the mean occupation number is given by $\bar n=1/[\exp{[\hbar\Omega/(k_BT)]}-1]$, known as the Bose-Einstein distribution (with $k_B$ the Boltzmann constant)~\cite{Clerk2010}. For thermal energies large compared to the energy quantum ($k_B T \gg \hbar\Omega$), the mean energy of the harmonic oscillator is $\bar E=k_BT$ in agreement with classical statistical mechanics, and $\hbar$ makes no appearance.  However, for zero temperature, the oscillator retains the zero-point energy $E_0=\hbar\Omega/2$, whose existence can be interpreted as a result of the finite value of Planck's constant.

A particularly striking experiment to demonstrate the 
existence of the quantum ground state of an oscillator is Raman scattering, where  light at the angular frequency $\omega$ is scattered into a Stokes sideband at $\omega-\Omega$ and an anti-Stokes sideband at $\omega+\Omega$. Stokes scattering is an inelastic process raising the population of the mechanical oscillator by a single quantum of energy (termed phonon), while anti-Stokes scattering corresponds to lowering the oscillator's population by one quantum. Importantly, anti-Stokes scattering is impossible by an oscillator in its quantum ground state. As a result, the powers in the anti-Stokes and Stokes sidebands differ. In the limit of $\Omega \ll \omega$, their ratio is given by $\bar n/(\bar n+1)=\exp{[-\hbar\Omega/(k_BT)]}$ and can serve as a temperature measurement calibrated relative to the quantum of energy of the system $\hbar\Omega$~\cite{Clerk2010}. 
In molecular systems, the oscillator frequency $\Omega$ can be sufficiently high to make the Raman-sideband asymmetry a feature of quantum mechanics routinely exploited even at room temperature~\cite{Kip1990,Cui1998,Ukil2012}. Furthermore, pioneering experiments using laser-cooling techniques have investigated atoms and atom-clouds in their motional ground states in optical traps~\cite{Diedrich1989,Jessen1992,Monroe1995}. 

During the last decades, quantum mechanics has been tested on increasingly massive objects~\cite{OConnell2010}. In particular, macroscopic mechanical oscillators are now being used for optical measurements operating at the limits set by quantum theory~\cite{Braginskii1977,Caves1980,Braginsky1992}. 
%A prominent application driving these developments is the detection of gravitational waves in an optical interferometer~\cite{Abbott2016}.
Together with the remarkable progress in measurement precision, optical techniques have been developed to not only sense but also control mechanical motion at the quantum level~\cite{Safavi-Naeini2012,Weinstein2014,Purdy2015,Underwood2015, sudhir2017}. 
Using the forces of light, nano- and micro-mechanical oscillators have been cooled to their quantum ground states in schemes relying both on autonomous~\cite{Teufel2011,Chan2011} and active-feedback mechanisms~\cite{Rossi2018}. Thus far, besides requiring cryogenic precooling, all experiments demonstrating optical quantum-control of mesoscopic mechanical oscillators rely on coupling the mechanical degree of freedom to an optical resonator to boost the light-matter interaction strength~\cite{Aspelmeyer2014,khalili2012}.

In this work, we transition a mesoscopic mechanical oscillator from the classical to the quantum domain without the need for cryogenic cooling nor requiring coupling to an optical cavity. The oscillator is a dielectric sphere with a diameter of \SI{136}{nm}, levitated in ultra-high vacuum in a single-beam optical dipole trap~\cite{Ashkin1977,Chang2010,Romero-Isart2011,Gieseler2012,Jain2016}. We use measurement-based linear-feedback cooling to reduce the effective temperature of the particle's center-of-mass motion from room temperature by seven orders of magnitude to observe the emergence of the Raman-sideband asymmetry in 
%\textcolor{blue}{a heterodyne detection of}
the light scattered by the particle. Sideband thermometry yields a phonon occupation number of $\bar n= 4$. 
%Our experiments put optically levitated systems at the forefront of optomechanical technologies to test quantum physics in experiments inaccessible by mechanically tethered oscillators. 

\paragraph{Experimental.}
    \begin{figure}[tb]
    \includegraphics[width=\figwidth]{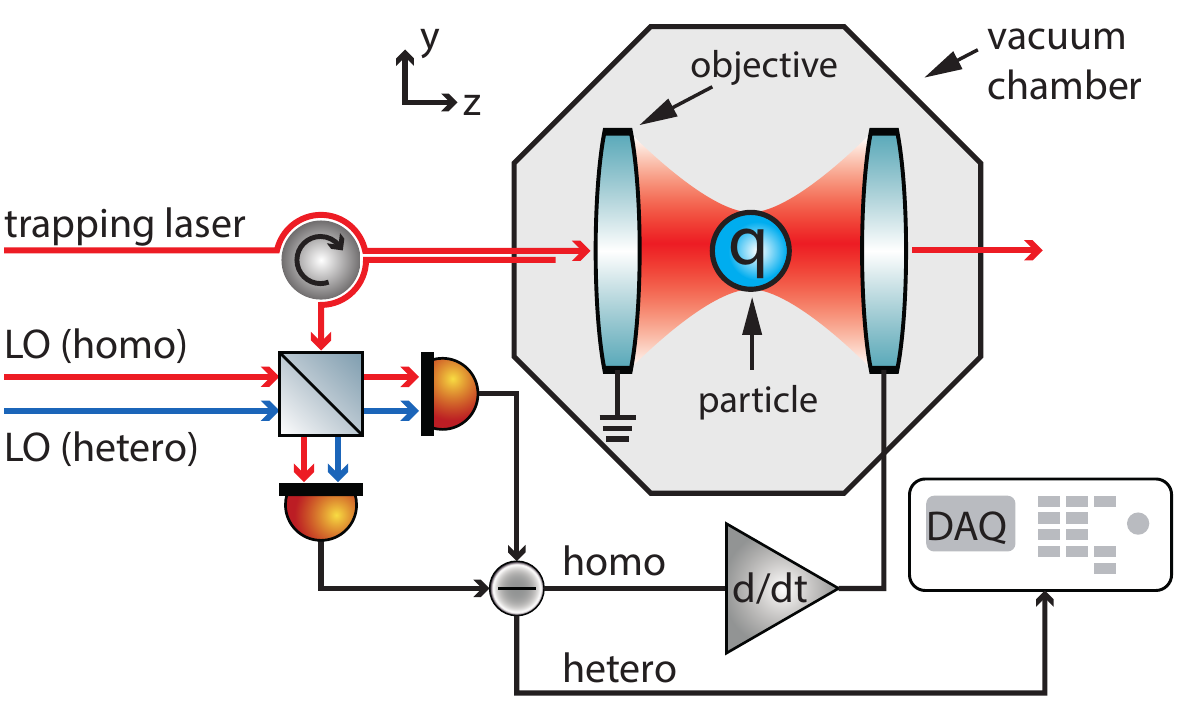}
    \caption{Experimental setup. A silica nanoparticle carrying a finite net charge $q$ is optically trapped in vacuum using a laser beam  focused by an objective. To measure the $z$ motion of the particle, the backscattered light is rerouted by a free-space circulator and mixed with two local oscillators (LO) for simultaneous homodyne (homo) and heterodyne (hetero) detection. The time derivative of the homodyne signal is applied to a capacitor enclosing the particle for cold damping. The heterodyne signal is recorded for sideband thermometry. 
    }
    \label{fig:setup}
    \end{figure}

Our experimental setup is shown in Fig.~\ref{fig:setup}. 
We focus a linearly polarized laser beam (wavelength \SI{1064}{nm}, focal power \SI{130}{mW}) with a microscope objective (\SI{0.85}{\NA}) in vacuum (\SI{7.5e-9}{mbar}) to generate an optical dipole trap for a silica nanoparticle (diameter \SI{136}{nm}). 
The oscillation frequencies of the particle's center-of-mass are $\Omega_{z} = 2\pi\times50~{\rm kHz}$, $\Omega_{x} = 2\pi\times130~{\rm kHz}$, and $\Omega_{y} = 2\pi\times150~{\rm kHz}$, where $z$ denotes the direction along the optical axis and $x$ ($y$) the coordinate in the focal plane along (orthogonal to) the axis of polarization.   
By means of parametric feedback, we cool the particle motion along the $x$ and $y$ directions to temperatures below \SI{1}{K} to eliminate non-linear cross-coupling between the translational degrees of freedom~\cite{Jain2016}. 
In the following we focus on the particle's motion along the optical $z$ axis.

To profit from a maximized measurement efficiency, we detect the motion of the particle along the $z$ axis using the light scattered back into the trapping objective~\cite{Tebbenjohanns2019Efficiency}. The backscattered light is sent through a Faraday rotator and detected in a balanced detection scheme. Here, we mix the signal beam with both a homodyne and a heterodyne (shifted by \SI{\pm1}{MHz}) reference beam. 
We refer to the homodyne backscattering measurement as the in-loop signal, since we use it to derive a feedback signal proportional to the particle's velocity $\dot z$ along the optical axis~\cite{Tebbenjohanns2019}. This feedback signal is applied as a voltage to a capacitor enclosing the trapped particle. The particle carries a finite net charge, such that the feedback signal directly translates into the Coulomb force $F_\text{fb}=-m\gamma_\text{fb}\dot z$ acting on the particle, with feedback gain $\gamma_\text{fb}$ and mass $m$. 
The heterodyne signal measured in backscattering is used for an out-of-loop measurement of the particle motion. It provides a simultaneous measurement of the Stokes and anti-Stokes sidebands and therefore allows for sideband thermometry~\cite{Purdy2015}. 

\paragraph{Results.}
        \begin{figure*}
        %\includegraphics[width=0.9\textwidth]{Fig2_Tebbenjohanns.pdf}
 %       \setlength{\unitlength}{0.1\textwidth}
 %       \begin{picture}(10,3.5)
 %            \put(0,0){\includegraphics[width=0.9\textwidth]{Fig2_Tebbenjohanns.pdf}}
 %            \put(5.3,0.8){$\bar n /(\bar n+1) = \sqrt{R_-/R_+} $}
  %           \put(5.3,1.2){\textcolor{red}{$\bar n +1= c \int \diff f~ \tilde S_{zz}^\text{het, l} (f)$}}
   %     \end{picture}
        \includegraphics[width=0.9\textwidth]{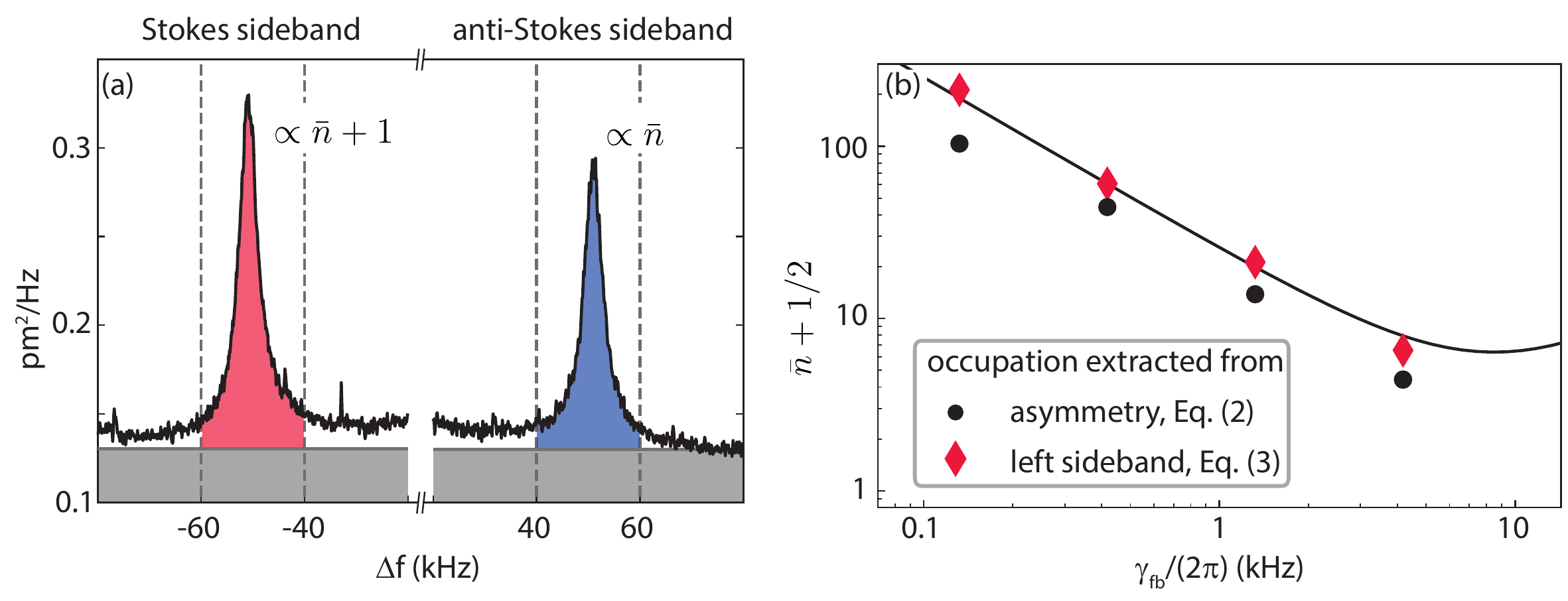}
        \caption{
        (a)~Motional sideband asymmetry. The figure shows single-sided power spectral densities $\tilde{S}^\text{het}_{zz}(f)$ obtained by the heterodyne out-of-loop measurement. The frequency difference $\Delta f$ is measured relative to the (absolute) local-oscillator frequency shift of \SI{1}{MHz}. Spectra are taken simultaneously under linear feedback cooling with $\gamma_\text{fb} = 2\pi \times \SI{4}{kHz}$. We observe an asymmetry in the power contained in the two sidebands. 
        The grey solid lines indicate the noise floor (limited by technical laser noise).
        The vertical dashed lines indicate the integration range.
        The calibration of the signal to absolute units follows the procedure outlined in Ref.~\cite{Hebestreit2018}.
        (b)~Mean occupation number as a function of feedback gain. The red diamonds are obtained by integrating the left sideband of the heterodyne spectrum according to Eq.~\eqref{eq:trad_calib}. 
        The black circles show the mean occupation number extracted from the sideband asymmetry according to Eq.~\eqref{eq:n/n+1}. The black solid line corresponds to a parameter-free model according to Ref.~\onlinecite{Tebbenjohanns2019}. Error bars (one standard deviation) are smaller than the symbol size.
        }
        \label{fig:e_vs_gain}
        \end{figure*}

In Fig.~\ref{fig:e_vs_gain}(a), we show the heterodyne sidebands generated by the motion of the particle along the $z$ axis. A feedback gain of $\gamma_\text{fb}=2\pi \times \SI{4}{kHz}$ is used and the local-oscillator frequency is shifted by \SI{-1}{MHz} relative to the trap laser. Each sideband has the shape of a Lorentzian function on top of an approximately constant noise floor. We observe that the left sideband at a frequency $\Delta f=\SI{-50}{kHz}$ (corresponding to Stokes scattering) carries more power than the right sideband at $\Delta f=+\SI{50}{kHz}$ (corresponding to anti-Stokes scattering). 
The power difference corresponds to the phonon energy $\hbar \Omega_z$ of the oscillator~\cite{Clerk2010,Safavi-Naeini2012,Jayich2012,Weinstein2014,Purdy2015}. As a result, the mean occupation number $\bar n$ is related to the sideband asymmetry
\begin{equation}\label{eq:R_-}
    R_- =  \frac{\int\diff f~ \tilde S_{zz}^\text{het,r}(f)}{\int\diff f~ \tilde S_{zz}^\text{het,l}(f)},
\end{equation}
where $\tilde S_{zz}^\text{het,r}$ ($\tilde S_{zz}^\text{het,l}$) is the power spectral density of the right (left) sideband. We derive $R_-$ from the measured power spectral densities shown in Fig.~\ref{fig:e_vs_gain}(a)~\footnote{We define our power spectral densities according to $\Braket{z^2} = \int_0^{\infty} {\rm d}f ~\tilde{S}_{zz}(f) = \int_{-\infty}^{\infty} {\rm d}\Omega ~ S_{zz}(\Omega)$}. The integration range used throughout this work is indicated by the grey vertical dashed lines, and the horizontal grey solid line shows the noise floor that is subtracted before integration of the signal.
We note that the measurement imprecision is not limited by quantum shot noise but by technical noise of the laser source. Importantly, the ratio $R_-$ can be influenced by the frequency-dependent transfer function of the measurement system. The measured asymmetry hence is $R_-=R_\text{TF} ~\bar n/(\bar n+1) $, where $R_\text{TF}$ is the ratio of the transfer function at the two sidebands. To eliminate this classical effect as a possible source for the sideband asymmetry, we swap the position of the left and the right sideband by switching the frequency shift of the heterodyne reference from $\SI{-1}{MHz}$ to $+\SI{1}{MHz}$. With this reversed frequency shift, the left (right) sideband corresponds to anti-Stokes (Stokes) scattering, and we determine the corresponding sideband asymmetry $R_+=R_\text{TF} ~(\bar n+1)/\bar n$. Based on the sideband asymmetries $R_+$ and $R_-$, we can extract the mean phonon occupation $\bar n$ from the relation~\cite{Clerk2010}
\begin{equation}\label{eq:n/n+1}
    \sqrt{\frac{R_-}{R_+}} = \frac{\bar n}{\bar n + 1}.
\end{equation}
In Fig.~\ref{fig:e_vs_gain}(b), we plot as black circles the mean occupation number $\bar n$ of the $z$ mode of the particle as deduced from the sideband asymmetry according to Eq.~\eqref{eq:n/n+1} as a function of feedback gain $\gamma_\text{fb}$. At a feedback gain of $\gamma_\text{fb}=2\pi\times\SI{4}{kHz}$, we obtain an occupation of $\bar n=4$.

In the following, we provide two cross-checks to corroborate our sideband-thermometry measurements. 
As a first check, we directly relate the power in the Stokes sideband to the energy of the motion, as commonly done in levitated optomechanics, using the relation
\begin{equation}\label{eq:trad_calib}
\bar n+1 = c \int \diff f ~ \tilde S_{zz}^\text{het, l}(f).
\end{equation}
The resulting phonon occupation is shown as red diamonds in Fig.~\ref{fig:e_vs_gain}(b).
Importantly, this procedure relies on a calibration factor $c$ which is determined in the mildly under-damped regime at \SI{10}{mbar}, where the particle is equilibrated to room temperature and behaves entirely classically~\cite{Hebestreit2018}. 
Therefore, the red diamonds in Fig.~\ref{fig:e_vs_gain}(b) can be interpreted as an energy measurement relative to the classical quantity $k_B T$ (with $T\sim\SI{300}{K}$). 
In contrast, the black circles in Fig.~\ref{fig:e_vs_gain}(b) represent a measurement relative to the quantum of energy $\hbar\Omega_z$.
The agreement between the two methods is satisfying. The observed difference can be ascribed to a systematic error of the classical calibration constant $c$, which is known to drift when reducing the pressure in the vacuum chamber~\cite{Hebestreit2018}. We note that we have  excluded any influence of (classical) laser intensity noise on the asymmetry exceeding the statistical uncertainty. To this end, we have compared sideband-thermometry measurements at different levels of laser intensity noise in the trap~\cite{sudhir2017}.

As a second consistency check, we compare our experimental results to the model of a cold-damped oscillator~\cite{Poggio2007}, following the procedure outlined in Ref.~\onlinecite{Tebbenjohanns2019}. To this end, we quantify the coupling of the particle to the thermal bath by performing ring-down and reheating experiments.
%As a second consistency check, we perform ring-down and reheating experiments at a low feedback strength ($\gamma_\text{fb} = 2\pi \times \SI{0.1}{kHz}$) to quantify the coupling of the particle to the thermal bath~\cite{Tebbenjohanns2019}. 
Together with the noise floor of the in-loop measurement, we obtain a parameter-free calculation of the expected energy under feedback cooling (black line in Fig.~\ref{fig:e_vs_gain}). The model (which relies on the classical energy-calibration constant) is in excellent agreement with the classically obtained measurements (red diamonds).

\paragraph{Discussion and conclusion.}
We have carried out two different measurements of the center-of-mass energy of a levitated oscillator. First, we have measured the energy relative to room temperature [red diamonds in Fig.~\ref{fig:e_vs_gain}(b)]. Second, we have measured the energy relative to the ground state energy $\hbar\Omega_z/2$ (black circles) and found satisfactory agreement between both methods.
Thus, our experiments bring an optically levitated oscillator from the classical to the quantum regime, where zero-point fluctuations have a sizable contribution to the particle's energy. 

Let us discuss the limits of our cooling experiments. Detection of the oscillation along the optical axis ($z$ mode) in backscattering should allow the phonon population to be cooled below unity~\cite{Tebbenjohanns2019Efficiency}. A straightforward route towards reaching this limit is to reduce laser noise on the detector to the shot noise limit in combination with a reduction in pressure by an order of magnitude to eliminate gas heating.

In conclusion, we have measured the sideband asymmetry in the motional spectrum of a levitated oscillator. 
This asymmetry is an unambiguous signature of the quantum ground state of the harmonic oscillator and arises in the limit of small phonon occupation numbers.
Using active feedback cooling, we have compressed the center-of-mass energy of a harmonic oscillator by more than seven orders of magnitude, transitioning the system from the classical realm to the quantum regime. Importantly, all previous demonstrations of cooling a mechanical oscillator to the quantum regime relied on cryogenic precooling and were accompanied by coupling to an optical cavity,
either in order to capitalize on autonomous resolved sideband cooling, or to boost the measurement efficiency in an active feedback cooling scheme. In contrast, we use a single laser beam to trap a nanoparticle in free space. This configuration requires little experimental overhead and offers the advantage of largely unobstructed measurements and the opportunity to control the trapping potential spatially and temporally via the light field.
These features of optically levitated oscillators hold promise for fundamental tests of physics in yet unexplored  parameter regimes~\cite{Geraci2010,Arvanitaki2013}. At the same time, the absence of an optical resonator removes any timing constraints posed by the finite response time of a cavity. This fact might prove beneficial for optomechanical control schemes relying on fast pulse sequences~\cite{Vanner2011}. 

\begin{acknowledgments}
This research was supported by ERC-QMES (Grant No.\ 338763), the NCCR-QSIT program (Grant No.\ 51NF40-160591), and SNF (Grant No.\ 200021L-169319/1). We thank M.~Aspelmeyer, T.~Donner, J.~Gieseler, F.~van der Laan, A.~Militaru, R.~Reimann, O.~Romero-Isart, and I.~Shomroni for valuable input and discussions.
\end{acknowledgments}

\bibliography{Literature_Tebbenjohanns_SidebandAsymmetry}

\end{document}